\documentclass[preprint,aps,showpacs,preprintnumbers]{revtex4}
\usepackage{amssymb}

\usepackage{epsfig}
\usepackage{dcolumn}
\usepackage{bm}

\begin{document}

\title{Variable-Range Hopping of Spin Polarons:\\
Magnetoresistance in a Modified Mott Regime}
\author{M. Foygel$^1$, R. D. Morris$^2$, and A. G. Petukhov$^{1,3}$}
\affiliation{$^1$Physics Department, South Dakota School of Mines and
Technology, Rapid City, SD 57701\\
$^2$USRA/RIACS, NASA Ames Research Center, MS 269-2, Moffet Field, CA 94035\\
$^3$Center for Computational Materials Science, Naval Research Laboratory,
Washington, DC 20375}

\date{\today}

\begin{abstract}
We analize electrical conductivity controlled by hopping of bound spin
polarons in disordered solids with wide distributions of electron energies
and polaron shifts (barriers). By means of percolation theory and Monte
Carlo simulations we have shown that in such materials at low temperatures,
when hopping occurs in the vicinity of the Fermi level, a hard polaron gap
does not manifest itself in the transport properties. This happens because as
temperature decreases the hopping polaron trades the decreasing electron and
polaron barriers for increasing hopping distance. As a result, in the
absence of the Coulomb correlation effects, in this variable-range
variable-barrier hopping regime, 
the electrical resistivity, $\rho $, as a function
of temperature, $T$, obeys a non-activation law: $\ln \left( \rho /\rho
_{0}\right) =\left( \widetilde{T}/T\right) ^{p}$ with $p=2/(d+2)$, where $d$
is the dimensionality of the system. It differs from the standard Mott law
for which $p=1/(d+1)$. Also, we studied the effects of upper and lower
boundaries in the polaron shift distribution on hopping conduction, which
may result in a partial re-entrance of the hard polaron gap. We discuss
possible applications to the problem of giant negative magnetoresistance in
dilute magnetic semiconductors and nanocomposites where for paramanetic
materials $p=3/(d+2)$.

\end{abstract}

\maketitle

\section{Introduction}

\label{intro} Electrical conductivity due to variable-range hopping (VRH)
in doped crystalline and amorphous semiconductors has been a
subject of intensive experimental and theoretical studies for more than
three decades \cite{MottDavis,SEbook, BottgerBryksin, Pollak}. The
temperature dependence of the VRH conductivity is described by stretched
exponents $\sigma \propto \exp \left( -T/T_{0}\right) ^{p}$ and depending on
the value of $p$ can be attributed to one of the two major regimes such as
the Mott ($p=1/4$) and the Shklovskii-Efros ($p=1/2$) regimes \cite
{MottDavis,SEbook}. Several observations of giant and colossal negative VRH 
\emph{magnetoresistance} in systems with strong spin disorder such as dilute
magnetic semiconductors (DMS) have been reported recently \cite{VanEsch,
Kreutz, Crowell}. It seems rather natural to attribute VRH magnetoresistance
to spin polarons that are believed to be a major cause of negative
magnetoresistance in DMS \cite{Kasuya, DietlSpalek, PF2000}. However, a
theory of VRH magnetoresistance has never been addressed in the literature,
partially, because of the belief that spin-polaron hopping may lead only to
a simple activation dependence of the conductivity which mimics
nearest-neighbor hopping \cite{Ioselevich}. We will demonstrate that in the
systems with significant disorder of the magnetic energies spin-polaron
hopping is described by a stretched exponent similar to that of the Mott
behavior.

The study of electrical conductivity in solids, which is controlled by
polaron hopping, has centered so far on materials where the polaronic part
of the activation energy is fixed \cite{MottDavis,Holstein,BottgerBryksin,
Emin,NP,Ioselevich,PF2000}. Indeed, the lowering of the free energy of the
system due to atomic spin alignment \cite{DietlSpalek} ( or lattice
polarization \cite{Emin} ) in the vicinity of the trapped carrier, the so
called polaron shift, and its associated hopping barrier are generally
determined by the spatial extent of the localized states between which the
hopping occurs. This is usually true no matter how strong the coupling
between a localized carrier and magnons (or phonons) and how many of them
are emitted or absorbed during the hopping event \cite{Emin1}. (Sometimes,
non-linear polaron effects may lead to an additional self-localization of
the bound state resulting in the actual collapse of its wave function \cite
{DietlSpalek} . These situations are out of the scope of the present paper
for they would hardly be favorable for the hopping process.) That is why in
standard lightly doped semiconductors the polaron hopping barrier can be
considered fixed \cite{DietlSpalek}. (This is also correct in heavily doped
and compensated nonmagnetic semiconductors, where at low temperatures
conductivity is controlled by electron hopping between aggregations of
impurities of a fixed size close to the screening length \cite{SEbook}.) In
all these cases, the bound polaron spectrum has a hard gap resulting in a
predominantly activation temperature dependence of conductivity even in the
variable-range hopping (VRH) regime \cite{BottgerBryksin,Ioselevich}.

Recent technological developments in producing novel dilute magnetic
semiconductors, especially $GaAs:Mn$, and magnetic nanostructures, such as $%
GaAs/ErAs,$ made it possible to observe electrical conductivity due to
hopping via magnetic impurities \cite{VanEsch} and nano-islands \cite{PRL99}
thus revealing a rich interplay of magnetic and electrical transport
properties. For instance, $Mn$ atoms play a dual role as acceptor sites and
magnetic ions in $GaAs$ resulting in the very unusual magnetic properties of
this material \cite{VanEsch,Bhatt}. Giant negative magnetoresistance has
been observed \cite{VanEsch,Kreutz,Crowell} below the metal-insulator
transition point of $GaAs:Mn$, which clearly displays the VRH behavior with
no sign of the hard magneto-polaron gap. The absence of this gap in the
electrical conduction experiments can be understood by taking into account
local fluctuations of the bound polaron shifts.  
These, in turn,
may be related to the distribution of exchange couplings between $Mn$ ions and
holes as  was suggested in Ref. \cite{Bhatt}. In addition, it is quite
possible that the fluctuations of the local spin polaron shifts are not
correlated with those of the ''bare'' electron energies. In heavily doped
and compensated samples \cite{SEbook} of $GaAs:Mn$, the former can be
determined by the total number of magnetic impurities in the localization
volume while the latter may depend on the electrically active ones. For $%
ErAs $ nanoislands embedded in $GaAs$ matrix \cite{PRL99} , the distribution
of polaron shifts follows the distribution of the islands' volumes.
Meanwhile, the ''bare'' electron spectrum of the islands is apparently
determined by the distribution of the islands sizes and shapes. We will show
that fluctuations of the local (lattice and spin) polaron shifts in certain
circumstances may wash out the hard polaron gap, leading to a substantial
modification of the standard Mott law for the electrical conductivity \cite
{Mott} even in the absence of  Coulomb correlation effects.

Bound spin (magnetic) polarons, in contrast to their lattice counterparts,
possess polaron shifts that, as well as their distribution, can be tuned by
applied magnetic field thus resulting in giant or even colossal
magnetoresistance \cite{NP,Ioselevich,PF2000,PRL99}. Externally
controlled switching off and on of the magneto-polaron effect also allows one
to better identify the hopping mechanism in magnetic materials.

The main goal of the present paper is to develop a consistent theoretical
approach to the description of variable-range hopping of polarons in materials
where disorder gives rise not only to  the electron-energy distribution but,
also, to the distribution of local polaron shifts. This approach is based on 
percolation theory and Monte Carlo simulations. In Section \ref{qual}, we
will present simple qualitative arguments concerning the main results of the
present paper. A rigorous theoretical derivation of these results is given
in Section \ref{resistivity}. Section \ref{montecarlo} deals with the Monte
Carlo simulations of the polaron hopping problem aimed at finding the universal
constants and functions that appear in the applications of the percolation
theory. Effects of the low boundary in the distribution function of the
polaron barriers will be considered in Section \ref{lowtemp}. In Section \ref
{gmr} we briefly discuss giant negative magnetoresistance in magnetic
semiconductors and nanostructures, a phenomenon that in some materials can
be controlled by variable-range variable barrier hopping of the spin
polarons.

\section{Qualitative Consideration}

\label{qual}

We will start from a qualitative description of the low-temperature
electrical resistivity in a system where localized states have 
electronic energies randomly distributed in the vicinity of the Fermi energy 
$E_{F}$. The derivation of the original Mott's law is based on the
assumption that the density of states in such a system near $E_{F}$ is
constant \cite{MottDavis,SEbook,Mott}. In addition to this assumption, we
will suppose that the states in questions possess randomly distributed
polaron barriers as well. Our goal is to find out how this factor would
modify the standard Mott's law for DC hopping conductivity.

In the original Mott's derivation, it is assumed that at a given temperature 
$T$ most of the hopping events involve  random sites with one-electron
energies belonging to a strip of width $\epsilon $ around the Fermi level.
In our case, we have to account for an additional random distribution of the
polaron barriers. Thus we generalize the Mott conjecture and introduce a
hyperstrip of width $\epsilon $ and height $W$, where $W$ is a scatter of
the polaron barriers. 
The probabilty of a typical hopping event for two sites belonging to this
hyperstrip reads: 
\begin{equation}
w(\epsilon ,W)\propto \exp \left( -\frac{2r}{a}-\frac{\epsilon +W/2}{T}%
\right) ,  \label{w_ij}
\end{equation}
where $a$ is the localization radius of the electron wave function, $T$ is
measured in the energy units, and $r$ is the average distance between the
sites of the hyperstrip: 
\begin{equation}
r\simeq N^{-1/d}=\left( G\epsilon W\right) ^{-1/d}.  \label{distance}
\end{equation}
Here $N$ is the concentration of the sites in the hyperstrip, $d=$2 or 3 is
the spatial dimension of the system and $G$ is a joint density of
(polaronic) states (DOS), such that $G\left( \epsilon _{0},W_{0}\right)
d\epsilon dW$ is a probability, per unit volume, to find a site with $%
\epsilon $ in the interval $\left( \epsilon _{0},\epsilon _{0}+d\epsilon
\right) $ and $W$ in the interval $\left( W_{0},W_{0}+dW\right) $. (As in
the original Mott's derivation \cite{MottDavis,Mott}, we assume that $G$ is
constant. However, this is not true if the Coulomb correlations are taken
into account \cite{SEbook}.)

After substituting (\ref{distance}) in (\ref{w_ij}), 
it is easy to find that the probability (\ref{w_ij}) as a function of $%
\epsilon $ and $W$ will reach its sharp maximum at 
\begin{equation}
\epsilon _{opt}=W_{opt}/2=\left( \frac{2T}{daG^{1/d}}\right)
^{d/(d+2)}\simeq T\left( \widetilde{T}_{0}^{(d)}/T\right) ^{2/(d+2)}.
\label{opt}
\end{equation}
The hopping conductivity, $\sigma $, is proportional to $w(\epsilon
_{opt},W_{opt})$. Therefore, substituting the optimal values (\ref{opt}) and
(\ref{distance})  in (\ref{w_ij}) yields the following expression for the
hopping resistivity:

\begin{equation}
\rho =1/\sigma =\rho _{0}\exp \left[ \left( \widetilde{T}_{0}^{(d)}/T\right)
^{2/(d+2)}\right] ,  \label{rho_opt}
\end{equation}
where

\begin{equation}
\widetilde{T}_{0}^{(d)}=\left( \frac{\beta _{0}^{(d)}}{Ga^{d}}\right) ^{1/2}
\label{T_tilda_d}
\end{equation}
and $\beta _{0}^{(d)}$ is a numerical factor of the order of unity.
Expression (\ref{rho_opt}) constitutes the main result of the present paper.
(In the next section, we will obtain this result by means of a more rigorous
percolation theory approach. Also, by using the Monte Carlo simulation
described in Section 4 we will be able to find the numerical factor $\beta
_{0}^{(d)}$ in Eq. (\ref{T_tilda_d}).) It can be seen that, in accordance
with the above derived formula, in the $3D$ ($2D$) case the well known
Mott's exponent 1/4 (1/3) (see \cite{SEbook,Mott}) should be substituted
with a somewhat larger exponent of 2/5 (1/2).

As the temperature goes up both the width and the height, (\ref{opt}), of
the optimal hyperstrip increase while the optimal hopping distance (\ref
{distance}) decreases. It means that with increasing temperature the
polaron chooses to jump over increasingly smaller distances at the cost of
overcoming the increasingly higher and higher polaron (and electron energy)
barriers. That is why we call this regime a variable-range, variable-barrier
hopping (VRVBH). In this regime, in contrast to the nearest-neighbor hopping
(NNH) of polarons, both the lattice \cite{MottDavis,BottgerBryksin} or the
magnetic ones \cite{PF2000} , the polaron hard gap is washed out by the
fluctuations of the polaron shifts (barriers).

If, however, there is a maximal polaron shift in a system, $W_{\max },$ then
at high enough temperatures, such that $T>T_{1}^{(d)}$, where

\begin{equation}
T_{1}^{(d)}=\left( \left( t_{0}^{(d)}W_{\max }/4\right)
^{d+1}/T_{M}^{(d)}\right) ^{1/d},  \label{T1}
\end{equation}
the optimal polaron shift (\ref{opt}) becomes than $W_{\max }.$ (Here $%
t_{0}^{(d)}$ is a numerical factor of the order of unity, which will be
determined later by  Monte Carlo simulation.) In this case, the
combined DOS, $G$, should be substituted with an ordinary electron DOS, $g,$
and the VRVBH law (\ref{rho_opt}) with a slightly modified Mott's law for
polaron VRH \cite{MottDavis,BottgerBryksin}

\begin{equation}
\rho =\rho _{0}\exp \left[ \left( T_{M}^{(d)}/T\right) ^{1/(d+1)}+\tau
_{d}W_{\max }/T\right] ,  \label{Mott}
\end{equation}
where the second (hard-gap activation) term under the exponent sign is small
compared to the first (Mott's) term. Here

\begin{equation}
T_{M}^{(d)}=\frac{\beta _{M}^{(d)}}{ga^{d}},  \label{T_M^d}
\end{equation}
where $\tau _{d}$ and $\beta _{M}^{(d)}$ are numerical factors of the order
of unity, which will be established later.

At low temperatures, both the electron, $g$, and the combined, $G$, DOS
cannot anymore be treated as constant in the vicinity of the Fermi level due
to the appearance of the soft Coulomb gap in the electron spectrum \cite
{SEbook}. The latter is related to the long-range Coulomb correlation
effects in the spatial distribution of electrons above and holes below the
Fermi energy. The maximal width of this gap can be estimated as \cite{SEbook}

\begin{equation}
\epsilon _{C}\simeq \left[ \left( e^{2}/\kappa \right) ^{d}g\right]
^{1/(d-1)}\simeq \left[ \left( e^{2}/\kappa a\right) ^{d}/T_{M}^{(d)}\right]
^{1/(d-1)},  \label{Coulomb_gap}
\end{equation}
where $\kappa $ is the dielectric constant of the material and $e$ is the
electron charge. This gap will modify the VRVBH law (\ref{rho_opt}) at low
temperatures when its width exceeds the width (\ref{opt}) of the optimal
strip: $\epsilon _{C}>\epsilon _{opt}$, i.e. when $T<T_{2}^{(d)}$ with

\begin{equation}
T_{2}^{(d)}\simeq \epsilon _{C}\left( \epsilon _{C}/\widetilde{T}%
_{0}^{(d)}\right) ^{2/d},  \label{T_2_d}
\end{equation}
where $\widetilde{T}_{0}^{(d)}$ is given by expression (\ref{T_tilda_d}%
). It can be seen that the temperature interval ($T_{2}^{(d)}<T<T_{1}^{(d)}$%
), where the VRVBH regime (\ref{rho_opt}) exists, may be sufficiently large
if the polaronic effect is strong enough. Based on the evaluations of $G$,
presented in Section 6, it is possible to show that the required condition $%
T_{2}^{(d)}<<T_{1}^{(d)}$ is equivalent to $W_{\max }>>\epsilon _{C}$.

\section{Calculation of Resistivity. A Percolation Theory Approach}

\label{resistivity}

Let us consider two sites (quantum dots, impurity centers or their
aggregations, etc.) $i$ and $j$\textit{\ }separated by a distance $\mathit{r}%
_{ij}$. To find the bound polaron hopping rate in the so called nonadiabatic
limit \cite{Holstein,Emin} one can employ the semiclassical approach
developed in \cite{Holstein,Emin} for the lattice small polarons (SP) or in 
\cite{PF2000} for the bound magnetic polarons (BMP). In both cases, however,
is possible to show (see Appendix) that if $i\rightarrow j$ tunneling
transition is governed either by strong multiphonon coupling (for the
lattice SP) or by thermodynamic fluctuations of the local magnetization (for
the BMP), when $W_{i}+W_{j}\geqslant \left| \widetilde{\epsilon }_{j}-%
\widetilde{\epsilon }_{i}\right| ,$ the hopping rate

\begin{equation}
\gamma _{ij}=\gamma _{ij}^{0}\exp \left[ -\frac{2\mathit{r}_{ij}}{a}-\frac{%
\left( W_{i}+W_{j}+\widetilde{\epsilon }_{j}-\widetilde{\epsilon }%
_{i}\right) ^{2}}{4\left( W_{i}+W_{j}\right) T}\right] ,  \label{gamma}
\end{equation}
where $a$ is the spatial-decay length of the localized wave function into
the semiconductor matrix, $\widetilde{\epsilon }_{l}=\epsilon _{l}\left(
0\right) -W_{l}$ is an equilibrium energy of the $l$-th site with the
localized carrier; $\epsilon _{l}\left( 0\right) $ is the ``bare'' electron
energy of the carrier-free state and $W_{l}$ is the bound polaron shift. The
latter describes the lowering of the free energy of the system due to either
lattice polarization or to atomic spin alignment in the vicinity of the
trapped carrier. (Usually $W_{l}$ is inversely proportional to the number of
polar or magnetic atoms in the localization volume \cite{Emin,PF2000}.)

If, however, electron-phonon (electron-magnon) coupling is not strong, i.e.
if $W_{i}+W_{j}<\left| \widetilde{\epsilon }_{j}-\widetilde{\epsilon }%
_{i}\right| ,$ the single-acoustic-phonon assisted tunneling of 
Miller-Abrahams type \cite{SEbook} prevails over the resonant one described
by Eq. (\ref{gamma}). Then (see \cite{PF2000,SEbook} for details)

\begin{equation}
\gamma _{ij}\propto \left( N_{P}+1/2\right) \pm 1/2,  \label{gamma1}
\end{equation}
where

\begin{equation}
N_{P}=\left[ \exp \left( \frac{\widetilde{\epsilon }_{j}-\widetilde{\epsilon 
}_{i}}{T}\right) -1\right] ^{-1}  \label{planck}
\end{equation}
is Planck's distribution function. Here, in Eq. (\ref{gamma}), ``+'' and
``-''correspond to the cases of phonon emission ($\widetilde{\epsilon }_{j}<%
\widetilde{\epsilon }_{i}$) and absorption ($\widetilde{\epsilon }_{j}>%
\widetilde{\epsilon }_{i}$), respectively.

The effective impedance between the localized sites in question \cite{SEbook}

\begin{equation}
Z_{ij}=\frac{T}{e^{2}\gamma _{ij}f_{i}\left( 1-f_{j}\right) }  \label{imp1}
\end{equation}
where $f_{l}=\left\{ 1+\exp \left[ \left( \widetilde{\epsilon }%
_{l}-E_{F}\right) /T\right] \right\} ^{-1}$ is Fermi's distribution function
of the bound polarons with Fermi level $E_{F}.$ From (\ref{gamma}) - (%
\ref{imp1}) it follows that

\begin{equation}
Z_{ij}=Z_{ji}\propto \exp \left( \frac{2r_{ij}}{a}+\frac{\epsilon
_{ij}+\Lambda _{ij}}{T}\right) ,  \label{imp2}
\end{equation}
where

\begin{equation}
\epsilon _{ij}=\frac{1}{2}\left( \left| \widetilde{\epsilon _{i}}%
-E_{F}\right| +\left| \widetilde{\epsilon _{j}}-E_{F}\right| +\left| 
\widetilde{\epsilon _{i}}-\widetilde{\epsilon _{j}}\right| \right)
\label{epsij}
\end{equation}
and

\begin{equation}
\Lambda _{ij}=\left\{ 
\begin{array}{c}
0,\;\;\;\qquad \qquad \qquad \qquad \qquad \qquad \qquad \left| \widetilde{%
\epsilon _{i}}-\widetilde{\epsilon _{j}}\right| >W_{i}+W_{j} \\ 
\left. \left( W_{i}+W_{j}-\left| \widetilde{\epsilon }_{j}-\widetilde{%
\epsilon }_{i}\right| \right) ^{2}/4\left( W_{i}+W_{j}\right) \right.
,\left| \widetilde{\epsilon _{i}}-\widetilde{\epsilon _{j}}\right| \leq
W_{i}+W_{j}
\end{array}
\right. .  \label{lambda}
\end{equation}
It should be mentioned that Eqs (\ref{imp2}) - (\ref{lambda}), though
resembling those previously obtained in \cite{Ioselevich,PF2000} , differ
from them in one important point. Here the modified BMP energies $\widetilde{%
\epsilon }_{l}=\epsilon _{l}\left( 0\right) -W_{l}$ play role of the
``bare'' electron energies $\epsilon _{l}\left( 0\right) $ used in \cite
{Ioselevich,PF2000}. In the absence of fluctuations of polaron shifts ($%
W_{i}=W_{j}=W$), this substitution is irrelevant because the same constant
polaron shift $W$ is incorporated into $E_{F}$ \cite{DietlSpalek}. It is,
however, substantial if these fluctuations play a significant role as they
do in the case of variable-range hopping of the bound polarons.

Now we are in a position to calculate the electrical resistivity of the
system by using the percolation theory approach \cite{SEbook}. We start from
the following bonding criterion (see Eq. (\ref{imp2})

\begin{equation}
\xi _{ij}=\frac{2r_{ij}}{a}+\frac{\epsilon _{ij}+\Lambda _{ij}}{T}\leq \xi .
\label{conn}
\end{equation}
Let us introduce the maximum values of $\widetilde{\epsilon }$, $W$, and $r$%
, compatible with this criterion,

\begin{equation}
\widetilde{\epsilon }_{opt}=T\xi ,\qquad W_{opt}=4T\xi ,\qquad r_{opt}=a\xi
/2,  \label{max}
\end{equation}
as well as the following dimensionless variables:

\begin{equation}
\overrightarrow{s}_{l}=\frac{\overrightarrow{r}_{l}}{r_{opt}},\qquad \delta
_{l}=\frac{\widetilde{\epsilon }_{l}-E_{F}}{\widetilde{\epsilon }_{opt}}%
,\qquad w_{l}=\frac{W_{l}}{W_{opt}}.  \label{var}
\end{equation}
In terms of these variables the bonding criterion (\ref{conn}) can be
rewritten as

\begin{equation}
s_{ij}+\delta _{ij}+\lambda _{ij}\left( \varphi \right) \leq 1,
\label{conn1}
\end{equation}
where

\begin{equation}
s_{ij}=\left| \overrightarrow{s}_{i}-\overrightarrow{s}_{j}\right| ,
\label{sij}
\end{equation}

\begin{equation}
\delta _{ij}=\frac{1}{2}\left( \left| \delta _{i}\right| +\left| \delta
_{j}\right| +\left| \delta _{i}-\delta _{j}\right| \right) ,  \label{deltaij}
\end{equation}
and

\begin{equation}
\lambda _{ij}\left( \varphi \right) =\left\{ 
\begin{array}{c}
0,\qquad \qquad \qquad \qquad \qquad \qquad \qquad \qquad \qquad \varphi
\left| \delta _{i}-\delta _{j}\right| \geq 4\left( w_{i}+w_{j}\right)  \\ 
\left[ 4\left( w_{i}+w_{j}\right) -\varphi \left| \delta _{i}-\delta
_{j}\right| \right] ^{2}/16\varphi \left( w_{i}+w_{j}\right) ,\;\;\varphi
\left| \delta _{i}-\delta _{j}\right| <4\left( w_{i}+w_{j}\right) 
\end{array}
\right. .  \label{lambdaij}
\end{equation}
Here a dimensionless parameter $\varphi $ accounts for a possible existance
of an upper boundary, $W_{\max }$, in the distribution of the polaron shifts 
$W_{l}$. If, however, the optimal polaron shift  $W_{opt}=4T\xi <W_{\max }$,
this boundary does not play any role, and one should put $\varphi =1$ in (%
\ref{lambdaij}). This is true at low temperatures, $T<T_{1}^{(d)}$ (\ref{T1}%
), otherwise $\varphi >1$.

Now the following percolation problem can be formulated in the $d+$2
dimensional space. There are randomly distributed sites in this space with
the dimensionless concentration

\begin{equation}
\nu _{0}^{(d)}\left( \xi \right) =G\cdot 2\widetilde{\epsilon }_{opt}\cdot
W_{opt}\cdot r_{opt}^{d}=2^{3-d}Ga^{d}T^{2}\xi ^{d+2},  \label{conc}
\end{equation}
such that each of these sites, $l$, has two random parameters, $\delta _{l}$
and $w_{l}$, uniformly distributed inside the hyperstrip with $\left| \delta
_{l}\right| <1$ and $0<w_{l}<1.$ Given the bonding criterion (\ref{conn1})
with $\varphi =1$, one should numerically find a minimal (threshold)
dimensionless concentration, $\nu _{c0}^{(d)}=\nu _{c0}^{(d)}\left( \xi
_{c0}^{(d)}\right) $, at which the percolation onset takes place. If such a
concentration is found then, in accordance with Eq. (\ref{conc}), the
critical exponent

\begin{equation}
\xi _{c0}^{(d)}=\left( \frac{\nu _{c0}^{(d)}}{2^{3-d}Ga^{d}T^{2}}\right)
^{1/(d+2)}  \label{ksi0}
\end{equation}
and therefore the resistivity

\begin{equation}
\rho =\rho _{0}\exp \left( \xi _{c0}^{(d)}\right)  \label{rho1}
\end{equation}
is given by the previously obtained Eqs (\ref{rho_opt}) and (\ref{T_tilda_d}%
), where the dimensionless factor

\begin{equation}
\beta _{0}^{(d)}=2^{d-3}\nu _{c0}^{(d)}.  \label{beta_0}
\end{equation}

If there is a maximal polaron shift $W_{\max }$ then at high temperatures $%
T>T_{1}^{(d)}$(\ref{T1}), such that $W_{opt}=4T\xi _{c0}^{(d)}\geq W_{\max }$%
, all the polaron hopping barriers are accessible. In this case, $GW_{opt}=g$
and the dimensionless concentration (\ref{conc}) should be substituted by

\begin{equation}
\nu _{1}^{(d)}\left( \xi \right) =g\cdot 2\widetilde{\epsilon }_{opt}\cdot
r_{opt}^{d}=2^{1-d}ga^{d}T\xi ^{d+1}.  \label{conc1}
\end{equation}
Then given the bonding criterion (\ref{conn1}) one should find the threshold
critical concentration $\nu _{c1}^{(d)}\left( \varphi \right) =\nu
_{1}^{(d)}\left( \xi _{c1}^{(d)}\right) $ in the ($d$+1)-dimension space for
any dimensionless temperature $\varphi =4T\xi _{c1}^{(d)}/W_{\max }>1.$

As far as this concentration is found, the critical exponent can be
expressed as

\begin{equation}
\xi _{c1}^{(d)}\left( \varphi \right) =\left[ \frac{2^{d-1}\nu
_{c1}^{(d)}\left( \varphi \right) }{ga^{d}T}\right] ^{1/(d+1)}.
\label{ksic1}
\end{equation}
So, the parameter $\varphi >1$ can be found as a solution of the following
transcendental equation:

\begin{equation}
\varphi =t\left[ \frac{\nu _{c1}^{(d)}\left( \varphi \right) }{\nu
_{c1}^{(d)}\left( \varphi \rightarrow \infty \right) }\right] ^{1/(d+1)}.
\label{phi}
\end{equation}
Here a dimensionless parameter 
\begin{equation}
t=\frac{4\left( T^{d}T_{M}^{(d)}\right) ^{1/(d+1)}}{W_{\max }}\geq
t_{0}^{(d)}  \label{t}
\end{equation}
has been introduced where the Mott's parameter $T_{M}^{(d)}$ is given by Eq (%
\ref{T_M^d}) with

\begin{equation}
\beta _{M}^{(d)}=2^{d-1}\nu _{c1}^{(d)}\left( \infty \right) .
\label{beta_M}
\end{equation}
(The minimal value of the parameter (\ref{t}), $t_{0}^{(d)},$ which defines
the transition temperature $T_{1}^{(d)}$ (\ref{T1}), can be found from Eq (%
\ref{phi}) by setting there $\varphi =1$.)

For the high-temperature VRH regime ($T>T_{1}^{(d)}$), the resistivity can
be presented in the following form

\begin{equation}
\rho =\rho _{0}\exp \left( \xi _{c1}^{(d)}\right) =\rho _{0}\exp \left[
\left( T_{M}^{(d)}/T\right) ^{1/(d+1)}f_{1}^{(d)}\left( t\right) \right] ,
\label{rho2}
\end{equation}
It obeys a modified VRH Mott law with

\begin{equation}
f_{1}^{(d)}\left( t\right) =\varphi \left( t\right) /t  \label{f1d}
\end{equation}
being a universal function of the dimensionless parameter (\ref{t}), where
the function $\varphi (t)$ itself is a solution of the transcendental
equation (\ref{phi}). Function $\varphi \left( t\right) $ can be found given
the threshold concentration, $\nu _{c1}^{(d)}$, as a function of $\varphi >1$%
. In particular, from (\ref{conn1}) - (\ref{lambdaij}) it follows that due
to the fact that $\lambda _{ij}\left( \varphi \right) >0$ and is a dropping
function, $\nu _{c1}^{(d)}\left( \varphi \right) /\nu _{c1}^{(d)}\left(
\infty \right) >1$ and is a dropping function too. As a result, $%
f_{1}^{(d)}\left( t\right) >1$ is also a dropping function. At high
temperatures, such that $T>>T_{1}$, parameter $t>>1$. Then the solution of
Eq. (\ref{phi}) can be presented as $\varphi \left( t\right) \simeq t$ and
therefore $f_{1}\left( t\right) \simeq 1$ in this limit. In this case, the
resistivity is governed by the conventional Mott's law. (Also, from Eq (\ref
{lambdaij}) it directly follows that $\lambda _{ij}\rightarrow 0$ when $%
\varphi \rightarrow \infty .$ In this limit, the bonding criterion (\ref
{conn1}) coincides with that of the standard VRH problem \cite{SEbook}.)

It should be noted that the condition $T=T_{1}^{(d)}$(\ref{T1}) is
equivalent to $\xi _{c0}^{(d)}=\xi _{c1}^{(d)}$, i.e. the transition from
Eq. (\ref{rho_opt}) to Eq. (\ref{rho2}) takes place when the widths of the
corresponding optimal energy strips, $\widetilde{\epsilon }_{opt}$ are equal
to each other and, simultaneously, the optimal polaron shift $%
W_{opt}=W_{\max }$.

\section{Polaron Hopping. Monte Carlo Simulations}

\label{montecarlo}

A Monte Carlo simulation procedure has been developed and implemented,
which aimed at finding numerical factors $\beta _{0}^{(d)},$ $\beta
_{M}^{(d)}$ and $\tau _{d}$ in Eqs (\ref{rho_opt}) and (\ref{Mott}), as well
as the universal function $f_{1}^{(d)}\left( t\right) $ and parameter $%
t_{0}^{(d)}$ in Eqs (\ref{rho2}) and (\ref{T1}) . This procedure can be
described as follows. In a cube or square of a size $L$, $N$ randomly
distributed sites are planted with an average concentration $\nu
_{L}=N/L^{d} $, less than the critical one. Each of these sites, $l$, is
ascribed two random parameters, $\delta _{l}$ and $w_{l}$, such that $\left|
\delta _{l}\right| <1$ and $0<w_{l}<1.$ Any two sites, $i$ and $j,$ are
considered connected (bonded) if they satisfy the pairwise bonding criterion
(\ref{lambdaij}) with $\varphi \geq 1$, where $s_{ij}$ is the distance
between these sites measured in the same units as $L$. Two sites are said to
belong to the same cluster if there is a sequence of connections from the
first site, through other sites, to the second one. (Each cluster has its
own identification number.) To test for percolation, boundary layers of
thickness $b$ are introduced. \cite{SEbook,PikeSeager}. For a given $\varphi
\geq 1,$ the code checks if a percolation cluster has been established,
which has at least one site belonging to each of a pair of  opposite boundary layers. If it
has not, new sites are added
in small increments and the clusters updated until
percolation is detected at a critical (threshold) concentration $\nu
_{cL}\left( \varphi \right) $. Obviously, the latter depends on a given
realization of the percolation cluster, so the procedure was repeated from
50 to 100 times in order to find an average critical concentration $\langle
\nu _{cL}\left( \varphi \right) \rangle .$

Thus found critical concentrations for several increasing sizes $L$ have
been used to determine a thermodynamic limit ($L\rightarrow \infty $) of the
critical concentration, $\nu _{c}\left( \varphi \right) ,$ by using the
following power-law extrapolation:

\begin{equation}
\langle \nu _{cL}\left( \varphi \right) \rangle =\nu _{c}\left( \varphi
\right) +AL^{-\alpha },  \label{extrapolation}
\end{equation}
where $A$ and $\alpha $ are constants. (It has been found that the
distribution of $\nu _{cL}\left( \varphi \right) $ is Gaussian with
dispersion that tends to zero as the size of the system increases.)

\begin{figure}[tbp]
\epsfig{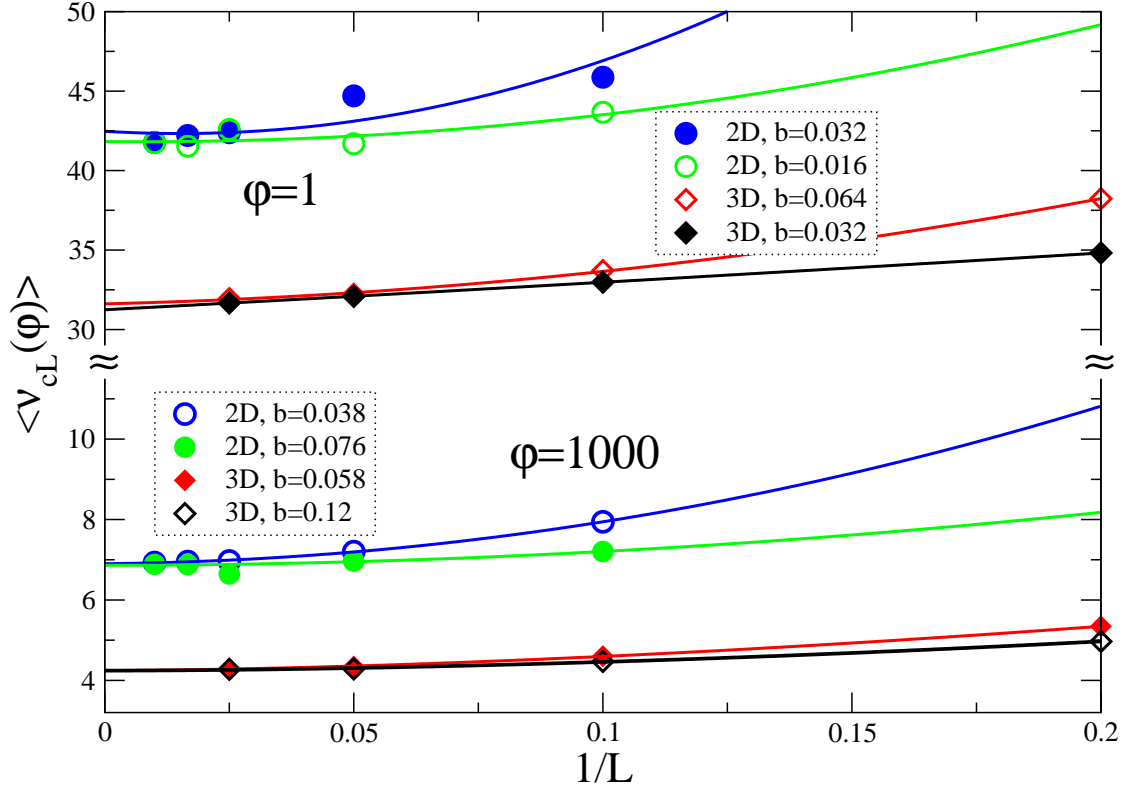}
\caption{ Average critical concentrations, $\langle \protect\nu _{cL}\left( 
\protect\varphi \right) \rangle $, as functions of size, $L$, of the $2D$
and $3D$ system calculated for  two different thicknesses, $b$, of the
boundary layer for $\protect\varphi =1$ (the upper four curves) and for $%
\protect\varphi =1000$ (the lower four curves). The fitted curves are
obtained by means of Eq. (\ref{extrapolation}) }
\label{fig1}
\end{figure}

Fig. \ref{fig1} clearly demonstrates that, for example, for $\varphi =1$ and 
$\varphi =1000$, the average critical concentrations as functions of the
size $L$ converge to their corresponding thermodynamic limits, $\nu
_{c}\left( \varphi \right) ,$ that do not depend on the thickness of the
boundary layer $b$. (In our simulations, the latter has been chosen to be
close to $0.2\nu _{c}^{-1/d}\left( \varphi \right) $ and $0.1\nu
_{c}^{-1/d}\left( \varphi \right) $ for any given $\varphi \geq 1$. And, as
has been explained in \cite{SEbook}, for any given $L,$ the thicker
boundary layer corresponds to a systematically smaller critical
concentration.)

In particular, we have found that the critical concentration $\nu
_{c}^{(d)}\left( 1\right) =\nu _{c0}^{(d)}$, that determines the critical
exponent (\ref{ksi0}) describing the conductivity (\ref{rho_opt}) in the
VBVRH regime, is equal to $42.19\pm 0.25$ and $31.25\pm 0.18$ for $2D$ and $%
3D$ systems, respectively. This yields the following values of the numerical
factor (\ref{beta_0}), which specify parameter $\widetilde{T}_{0}^{(d)}$ in
Eq (\ref{rho_opt}): $\beta _{0}^{(2)}=21.09\pm 0.12$ and $\beta
_{0}^{(3)}=31.25\pm 0.18.$

Also, the above described procedure has allowed us to establish values of
the critical concentrations in the limit of large $\varphi $: $\nu
_{c1}^{(2)}\left( \infty \right) =6.88\pm 0.08$ and $\nu _{c1}^{(3)}\left(
\infty \right) =4.24\pm 0.03$. These, in turn, determine the numerical
factors (\ref{beta_M}) in the Mott's law parameters (\ref{T_M^d}): $\beta
_{M}^{(2)}=13.76\pm 0.16$ and $\beta _{M}^{(3)}=16.96\pm 0.12.$ They can be
compared with previously obtained values $\beta _{M}^{(2)}=13.8\pm 1.2$ and $%
\beta _{M}^{(3)}=21.2\pm 1.2$ \cite{SEbook} (Our somewhat smaller, and more
accurate, value of $\beta _{M}^{(3)},$can be justified by using up to $%
3\times 10^{5}$ sites in the present simulations vs the previously
used $1.5\times 10^{2}$ sites \cite{SEbook}.) It should be mentioned that
our values of $\beta _{M}^{(d)}$ are in a surprisingly good agreement with
the ones ($\beta _{M}^{(2)}=13.3$ and $\tau _{M}^{(3)}=16.4$) obtained by
Ioselevich \cite{Ioselevich1} who applied an approximate multicomponent
percolation criterion to analysis of both the NNH and VRH problems.

\begin{figure}[tbp]
\epsfig{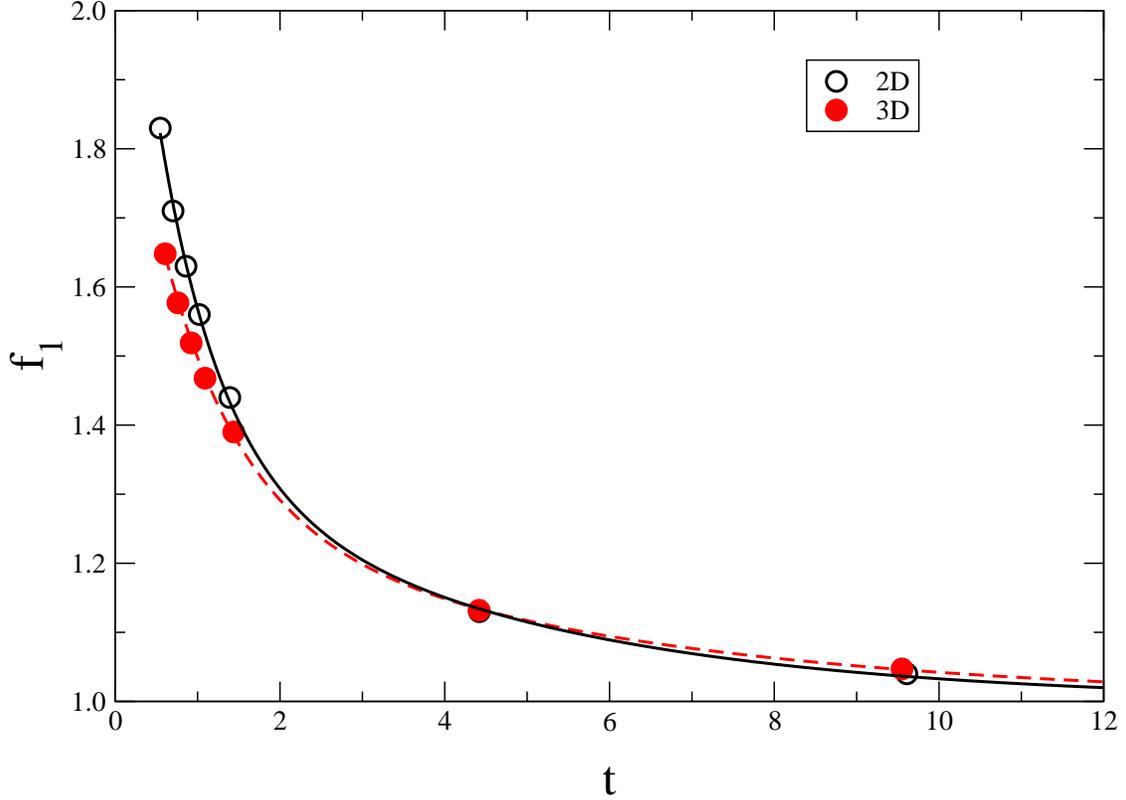}
\caption{ Universal function $f_{1}^{(d)}\left( t\right) $ found by means of
the Monte Carlo simulations (circles). The solid curves correspond to
extrapolation formula (\ref{f1approx}) }
\label{fig2}
\end{figure}

Fig. \ref{fig2} shows the calculated universal function (\ref{f1d}) which
describes a continuous transition from the VRVBH regime of polaron
conductivity (\ref{rho_opt}) to the standard Mott regime. For $t\geq t_{0}$,
it can be approximated as

\begin{equation}
f_{1}^{(d)}\left( t\right) =a+bt^{-1}+ct^{-2}.  \label{f1approx}
\end{equation}
Fitting parameters for $2D$ and $3D$ cases, as well as parameters $\tau
_{d}=b/4$ from Eq. (\ref{Mott}) and other relevant parameters can be found
in Table \ref{table1}.

\begin{table}[tbp]
\caption{Parameters of the universal function $f_{1}^{(d)}\left( t\right)$.}
\label{table1}%
\begin{ruledtabular}
\begin{tabular}{lccccccc}
& $a$ & $b$ & $c$ & $t_{0}$ & $\beta _{0}$ & $\beta _{M}$ & $\tau $ \\ 
\hline
$d=2$ & 0.967 & 0.756 & -0.157 & 0.546 & 21.1 & 13.8 & 0.189 \\ 
$d=3$ & 0.982 & 0.694 & -0.178 & 0.607 & 31.2 & 17.0 & 0.174 \\ 
\end{tabular}
\end{ruledtabular}
\end{table}

\section{Low Temperatures. Minimum Polaron Barrier Effect}

\label{lowtemp}

In certain disordered systems, such as doped semiconductors close to the
metal-insulator transition, the ditribution of polaron shifts, being broad
on its high side, may be very sharp on its low side \cite{Ioselevich}. It
means that a system may have some minimal polaron shift, $W_{\min }$. In
this case, the law (\ref{rho_opt}) is also violated at low temperatures when
the optimal polaron shift becomes smaller than the minimal one: $%
W_{opt}=4T\xi _{c0}^{(d)}<W_{\min }$, i.e. at $T<T_{3}$, where

\begin{equation}
T_{3}^{(d)}\simeq \left( W_{\min }/W_{\max }\right) ^{(d+1)/d}T_{1}^{(d)}.
\label{T3}
\end{equation}
where $T_{1}^{(d)}$ is given by Eq (\ref{T1}). (The regime described by Eq. (%
\ref{rho1}) can be observed within a temperature interval $%
max(T_{2}^{(d)},T_{3}^{(d)})<T<T_{1}^{(d)}$ that clearly exists if $\left(
W_{\max }-W_{\min }\right) /\epsilon _{C}>>1.$) At such low temperatures, $%
W_{\min }/2$ is the only polaron barrier accessible and the ``bare'', $%
\epsilon _{l}\left( 0\right) $ or the modified, $\widetilde{\epsilon }%
_{l}=\epsilon _{l}\left( 0\right) -W_{\min },$ electron energy is the only
fluctuating variable. The connectivity criterion (\ref{conn}) can be
rewritten as $\xi _{ij}=\xi _{ij}^{0}+\Delta \xi _{ij}<\xi +\Delta \xi _{ij}$%
, where $\Delta \xi _{ij}=W_{\min }/2T$ and

\begin{equation}
\xi _{ij}^{0}=\frac{2r_{ij}}{a}+\frac{\epsilon _{ij}+\Lambda _{ij}^{0}}{T}.
\label{conn2}
\end{equation}
Here

\begin{equation}
\Lambda _{ij}^{0}=\left\{ 
\begin{array}{c}
-W_{\min }/2,\;\;\;\qquad \qquad \qquad \qquad \qquad \qquad \left| 
\widetilde{\epsilon _{i}}-\widetilde{\epsilon _{j}}\right| >2W_{\min } \\ 
\left( \widetilde{\epsilon _{i}}-\widetilde{\epsilon _{j}}\right)
^{2}/8W_{\min }-\left| \widetilde{\epsilon _{i}}-\widetilde{\epsilon _{j}}%
\right| /2,\qquad \qquad \left| \widetilde{\epsilon _{i}}-\widetilde{%
\epsilon _{j}}\right| \leq 2W_{\min }
\end{array}
\right.   \label{lambdaij0}
\end{equation}
After introducing the maximum values of $\widetilde{\epsilon }_{l}$ and $%
\overrightarrow{r}_{l}$ (see (\ref{max}), compatible with the criterion $\xi
_{ij}^{0}<\xi $, as well as the dimensionless variables $\overrightarrow{s}%
_{l}$ and $\delta _{l}$ (Eq. (\ref{var})), the following percolation problem
can be formulated (see also \cite{Ioselevich}). For any given dimensionless
temperature, $\theta =2T\xi _{c2}^{(d)}/W_{\min }$, one should find the
critical dimensionless concentration, $\nu _{c2}^{(d)}\left( \theta \right) $%
, when percolation first appears given the connectivity criterion

\begin{equation}
s_{ij}+\delta _{ij}+\lambda _{ij}^{0}\left( \theta \right) \leq 1
\label{conn3}
\end{equation}
where

\begin{equation}
\lambda _{ij}^{0}\left( \theta \right) =\left\{ 
\begin{array}{c}
-1/\theta ,\qquad \qquad \qquad \qquad \qquad \qquad \theta \left| \delta
_{i}-\delta _{j}\right| \geq 4 \\ 
\theta \left( \delta _{i}-\delta _{j}\right) ^{2}/16-\left| \delta
_{i}-\delta _{j}\right| /2,\qquad \qquad \theta \left| \delta _{i}-\delta
_{j}\right| <4
\end{array}
\right. ,  \label{lambda_0}
\end{equation}
and assuming that $\delta _{l}$ are uniformly distributed within the strip $%
\left| \delta _{l}\right| <1.$ When such critical concentration is known,
then the critical exponent, $\xi _{c2}^{(d)}$, is given by Eq. (\ref{ksic1})
where $\nu _{c1}^{(d)}\left( \varphi \right) $ should be substituted by $\nu
_{c2}^{(d)}\left( \theta \right) .$ The resistivity can be presented in the
following form

\begin{equation}
\rho =\rho _{0}\exp \left[ \frac{W_{\min }}{2T}+\xi _{c2}^{(d)}\left( \theta
\right) \right] =\rho _{0}\exp \left[ \frac{W_{\min }}{2T}f_{2}^{(d)}\left(
\eta \right) \right] ,  \label{rho3}
\end{equation}
where 
\begin{equation}
\eta =\frac{2\left( T^{d}T_{M}^{(d)}\right) ^{1/(d+1)}}{W_{\min }}.
\label{eta1}
\end{equation}
Here $T_{M}^{(d)}$ is given by Eq. (\ref{T_M^d}) and $f_{2}^{(d)}\left( \eta
\right) =1+\theta \left( \eta \right) .$ Function $\theta \left( \eta
\right) $ is a solution of the transcendental equation

\begin{equation}
\theta =\eta \left[ \frac{\nu _{c2}^{(d)}\left( \theta \right) }{\nu
_{c2}^{(d)}\left( \theta \rightarrow \infty \right) }\right] ^{1/(d+1)},
\label{teta}
\end{equation}
where $\nu _{c2}^{(d)}\left( \theta \rightarrow \infty \right) =\nu
_{c1}^{(d)}\left( \varphi \rightarrow \infty \right) $ has been calculated
in the previous section.

As it follows from Eq. (\ref{lambda_0}), $\lambda _{ij}^{0}\left( \theta
\right) <0$ and is a rising function. As a result, $\nu _{c2}^{(d)}\left(
\theta \right) /\nu _{c2}^{(d)}\left( \infty \right) <1$ and is a rising
function. Therefore, in Eq. (\ref{rho3}), $f_{2}^{(d)}\left( \eta \right) $
is a rising function as well. In addition, if $\theta \rightarrow 0$, $\nu
_{c2}^{(d)}\left( \theta \right) /\nu _{c2}^{(d)}\left( \infty \right)
\rightarrow \varsigma _{d}<1$. Then at low temperatures , when $\eta $(\ref
{eta1})$<<1$, $f_{2}^{(d)}\left( \eta \right) \simeq 1+\varsigma
_{d}^{1/(d+1)}\eta $ which yields \cite{Ioselevich}

\begin{equation}
\rho =\rho _{0}\exp \left[ \frac{W_{\min }}{2T}+\left( \frac{\varsigma
_{d}T_{M}^{(d)}}{T}\right) ^{1/(d+1)}\right] .  \label{rho4}
\end{equation}
This almost activation type of conductivity (for the second term in the
exponent is much smaller than the first one) will be observed if $%
T<<T_{3}^{(d)}$.

The Monte Carlo simulations allowed us to calculate the universal function $%
f_{2}^{(d)}\left( \eta \right) $ in Eq. (\ref{rho3}), which can be
approximated as

\begin{equation}
f_{2}^{(d)}\left( \eta \right) =1+\eta \left[ 1-A\exp \left( -B\eta \right) %
\right] .  \label{f2}
\end{equation}
Fitting parameters for this formula and constant $\varsigma _{d}=\left(
1-A\right) ^{d+1}$ from Eq. (\ref{rho4}) are presented in Table \ref{table2}.

\begin{table}[tbp]
\caption{Parameters of the universal function $f_{2}^{(d)}\left(\protect\eta%
\right)$. }
\label{table2}%
\begin{ruledtabular}
\begin{tabular}{lccc}
& $A$ & $B$ & $\varsigma $ \\ \hline
$d=2$ & 0.137 & 0.0759 & 0.643 \\ 
$d=3$ & 0.109 & 0.0612 & 0.630 \\ 
\end{tabular}
\end{ruledtabular}
\end{table}

And finally, if $W_{\max }-W_{\min }$ is smaller than the width $\epsilon
_{opt}$ (\ref{opt}) of the Mott strip, the VRVBH regime, described by Eq. (%
\ref{rho_opt}) ceases to exist. The fluctuations of polaronic shifts are not
important anymore ($W_{\max }\simeq W_{\min }$), and, if the Coulomb
correlation effects are not substantial ($W_{\max }>>\epsilon _{C}$), the
resistivity is described by Eq. (\ref{rho3}) at all relevant temperatures.
It is interesting to mention that in the high-temperature case ($t\simeq
\eta >>1,$), when $f_{1}^{(d)}\left( t\right) \simeq 1$ and $%
f_{2}^{(d)}\left( \eta \right) \simeq \eta $, both formulas (\ref{rho2}) and
(\ref{rho3}) lead to practically the same modified Mott law.

\section{Application to Giant Magnetoresistance}

\label{gmr}

In the previous sections, we described a new mechanism of polaron
conductivity controlled by variable-range hopping which leads to a
non-activation temperature dependence

\begin{equation}
\rho =\rho _{0}\exp \left[ \left( \widetilde{T}/T\right) ^{p}\right]
\label{rho_p}
\end{equation}
with $p=2/(d+2)$. This result, which differs from the standard Mott's $%
p=1/(d+1)$, can be justified for both the lattice and the spin polarons if
their polaron shifts (barriers) fluctuate independently from their bare
electron energies. However, the case of bound spin (or magnetic) polarons is
of  special interest because not only the polaron shifts but their
distribution can be altered under the action of the external magnetic field, 
$H$. It can be shown, that, indeed, the combined density of polaron states, $%
G$, which defines parameter $\widetilde{T}$ in Eq. (\ref{rho_p}) (see Eqs (%
\ref{rho_opt}) and (\ref{T_tilda_d})), will decrease with magnetic field. As
a result, the resistivity will sharply (exponentially) decrease with
magnetic field, so that Eq. (\ref{rho_opt}) also describes giant (or
colossal) negative magnetoresistance observed in dilute magnetic
semiconductors and nanostructures. 

To demonstrate this, let us suppose that, for the sake of simplicity, the
``bare'' electron energies $\epsilon $ and the polaron shifts $W$ and are
not correlated and both are distributed uniformly, each in the interval $(%
\overline{\epsilon }-\Delta \epsilon ,\overline{\epsilon }+\Delta \epsilon )$
and $\left( W_{\min },W_{\max }\right) $, correspondingly. Then the density
of the ``bare'' electron states

\begin{equation}
g=\left\{ 
\begin{array}{c}
0,\qquad \qquad \qquad \left| \epsilon -\overline{\epsilon }\right|
\geqslant \Delta \epsilon \\ 
g_{0}=N_{t}/2\Delta \epsilon ,\;\;\left| \epsilon -\overline{\epsilon }%
\right| <\Delta \epsilon
\end{array}
\right. ,  \label{g0}
\end{equation}
while the density of the BMP shifts

\begin{equation}
g_{W}=\left\{ 
\begin{array}{c}
0,\;\;\qquad \left| W-\overline{W}\right| \geqslant \Delta W \\ 
N_{t}/2\Delta W,\;\left| W-\overline{W}\right| <\Delta W\;
\end{array}
\right. ,  \label{gW}
\end{equation}
where $N_{t}$ is the concentration of the sites and $2\Delta W=W_{\min
}-W_{\max },$ $\overline{W}=\left( W_{\min }+W_{\max }\right) /2$. For the
chosen distributions,

\begin{equation}
G\simeq \frac{N_{t}}{2\Delta \epsilon \cdot 2\Delta W}=\frac{g_{0}}{2\Delta W%
}=\frac{g_{0}g_{W}}{N_{t}}  \label{G}
\end{equation}
given $\Delta \widetilde{\epsilon }\simeq \Delta \epsilon >>\Delta W$.

For bound spin (magnetic) polarons, their shifts, $W$, proportional to the
magnetic susceptibility, $\chi \left( H,T\right) $ \cite{Ioselevich,
PF2000,DietlSpalek}. The same is true for the width, $2\Delta W=W_{\min
}-W_{\max }$, of the shift distribution (\ref{gW}).  
Therefore, by
aligning the atomic spins the applied magnetic field not only reduces the
polaron hopping barriers but also makes their distribution sharper. Then,
from Eqs. (\ref{T_tilda_d}) and (\ref{G}) it follows that parameter $%
\widetilde{T}_{0}^{(d)}$in expression (\ref{rho_opt})

\begin{equation}
\widetilde{T}_{0}^{(d)}(H)\propto \sqrt{\chi \left( H,T\right) }.
\label{T0(H)}
\end{equation}
It substantially decreases with magnetic field when the magnetization
approaches its saturation value thus leading to giant or colossal negative
magnetoresistance.

If the material is in the paramagnetic phase and if the anti-ferromagnetic
coupling between the magnetic atoms is suppressed, then the susceptibility
in Eq. (\ref{T0(H)}) obeys the Curie law: $\chi \left( H,T\right) \propto
1/T $. In this case, the exponent $p=2/(d+2)$ in expression (\ref{rho_p})
describing VRVBH regime should be substituted with a somewhat larger
quantity $p=3/(d+2)$. It can be seen that in this regime for $3D$ systems, $%
0.4<p<0.6.$ That kind of dependence can be easily confused with the standard 
$p=0.5$ typical for electron VRH in the presence of Coulomb correlations 
\cite{SEbook}. (The effect of these correlations on VRH of the spin and lattice
polarons will be considered elsewhere. In particular, it can be shown that
in this case $p=\left( d+1\right) /\left( 2d+1\right) $ or $p=\left(
d+2\right) /\left( 2d+1\right) $ in the paramagnetic phase.) However, at
high magnetic fields (for spin polarons) the $p=2/(d+2)$ or $p=3/(d+2)$
VRVBH dependence (for the spin polarons) will be replaced by the $p=1/(d+1)$
Mott's law while at the same conditions the presence of the soft Coulomb gap
will result in the $p=0.5$ dependence.

Giant negative magnetoresistance in the variable-range hopping region
apparently has been observed in recently discovered dilute magnetic
semiconductors, such as GaAs:Mn \cite{VanEsch,Kreutz,Crowell} and $MnGe$ 
\cite{Park} below the metal-insulator transition point. In all these
materials, magnetoresistance is described by Eq. (\ref{rho_p}) with $%
p=0.25\div 0.56$ and $\widetilde{T}$ significantly decreasing with magnetic
field. Additional, more detailed measurements of the transport and magnetic
properties in question are needed to more specifically identify and
quantitatively describe the hopping mechanism in these materials. Also, the
effect of the soft Coulomb gap should be taken into account when describing
electrical conductivity governed by variable-range polaron hopping at very
low temperatures.

\section{Conclusions}

\label{concl}

We have examined variable-range hopping of polarons in polar and/or magnetic
disordered materials. In addition to electron impurity band effects, where
site positions and the ''bare'' electron energies are randomly scattered, we
took into account that the polaron shifts may be widely distributed as well.
It results in a significant modification of the law that governs the
low-temperature electrical conductivity in such materials. In the absence of
 Coulomb correlation effects, the standard Mott dependence should be
substituted by another, non-activation one, given by Eq. (\ref{rho_opt}),
which describes the variable-range variable-barrier hopping of lattice or spin
polarons.

\begin{acknowledgments}
Useful discussions with P. A. Crowell are greatly acknowledged. This work
has been supported by NSF grant No. DMR-0071823.
\end{acknowledgments}

\appendix

\section{Polaron Hopping Rate}

Here we will derive Eq. (\ref{gamma}) for the polaron hopping rate. We will
use a semi-classical approach \cite{Holstein} to describe lattice polaron
hopping in the framework of a generalized molecular-crystal model \cite{Emin}%
. We will further generalize this model order to take into account that the
initial and final states may have different polaron shifts.

In the Holstein's occurrence probability approach, the polaron hopping rate
between the initial state $i$ and the final state $j$ can be written as \cite
{Emin,Emin1}

\begin{equation}
\gamma _{ij}=\int\limits_{-\infty }^{\infty }d\stackrel{.}{E}P\left( 
\stackrel{.}{E}\right) \left| \stackrel{.}{E}\right| \left\langle \delta
\left( \epsilon _{j}-\epsilon _{i}\right) \delta \left( \stackrel{.}{%
\epsilon }_{j}-\stackrel{.}{\epsilon }_{i}-\stackrel{.}{E}\right)
\right\rangle .  \label{gamma2}
\end{equation}
Here

\begin{equation}
\epsilon _{i}=\epsilon _{i}\left( 0\right)
-\sum\limits_{n=1}^{L_{i}}A_{n}^{(i)}x_{n};\qquad \epsilon _{j}=\epsilon
_{j}\left( 0\right) -\sum\limits_{n=1}^{L_{j}}A_{n}^{(j)}y_{n}
\label{epsilons}
\end{equation}
are the local electron energies of the initial an final states coupled to
local vibrational modes $x_{n}$ and $y_{n}$ with $A_{n}^{(i,j)}$ being the
coupling constants ($L_{l}$ is the total number of the vibration modes
coupled to the localized carrier at $l$-th site) and

\begin{equation}
\stackrel{.}{E}=\left( \stackrel{.}{\epsilon }_{j}-\stackrel{.}{\epsilon }%
_{i}\right) _{\epsilon _{j}=\epsilon _{i}}  \label{E_dot}
\end{equation}
being the time rate of change of the relative electronic energies evaluated
at the event when the electronic terms coincide. In Eq.(\ref{gamma2}), the
brackets indicate a thermal average over all the configurational coordinates 
$x_{n}$ and $y_{n}$ and the velocities $\stackrel{.}{x}_{n}$ and $\stackrel{.%
}{y}_{n}$; $P\left( \stackrel{.}{E}\right) $ is the probability that the
carrier will perform a hop at the coincidence event.

By using the integral representation of the $\delta $-functions and the
harmonic approximation for the vibrational energy (see Ref. \cite{Emin} for
details) one can perform averaging in Eq.(\ref{gamma2}) and rewrite it as

\begin{eqnarray}
\gamma _{ij} &=&\left[ 4\pi T\sqrt{\left( W_{i}\Omega _{i}^{2}+W_{j}\Omega
_{j}^{2}\right) \left( W_{i}+W_{j}\right) }\right] ^{-1}\exp \left\{ -\left[
\left( \Delta +2W_{i}\right) ^{2}/4\left( W_{i}+W_{j}\right) -J_{ij}\right]
/T\right\}  \nonumber \\
&&\times \int\limits_{-\infty }^{\infty }d\stackrel{.}{E}P\left( \stackrel{.%
}{E}\right) \left| \stackrel{.}{E}\right| \exp \left[ -\stackrel{.}{E}%
^{2}/4T\left( W_{i}\Omega _{i}^{2}+W_{j}\Omega _{j}^{2}\right) \right] ,
\label{gamma3}
\end{eqnarray}
where $\Delta =$ $\stackrel{.}{\epsilon }_{j}\left( 0\right) -\stackrel{.}{%
\epsilon }_{i}\left( 0\right) $ is the electron energy difference between
the carrier-free states; $J_{ij}=J_{0}\exp \left( -r_{ij}/a\right) $ is the
hopping integral;

\begin{equation}
W_{l}=\sum\limits_{n=1}^{L_{l}}\left( A_{n}^{(l)}\right)
^{2}/2k_{n}^{(l)};\qquad \Omega
_{l}^{2}=W_{l}^{-1}\sum\limits_{n=1}^{L_{l}}\left( \omega
_{n}^{(l)}A_{n}^{(l)}\right) ^{2}/2k_{n}^{(l)}  \label{W_l}
\end{equation}
are the polaron shift and the squared average vibrational frequency of the $%
l $-th site, correspondingly. Here $\omega _{n}^{(l)}$ and $k_{n}^{(l)}$ are
respectively the frequency and the stiffness constant of $n$-th vibrational
mode that belongs to site $l$.

In the so called non-adiabatic limit, when the hopping integral $%
J\rightarrow 0$, the probability \cite{Holstein,Emin}

\begin{equation}
P\left( \stackrel{.}{E}\right) =2\pi J_{ij}^{2}/\hbar \stackrel{.}{\stackrel{%
.}{E}<<1},  \label{P}
\end{equation}
and therefore

\begin{equation}
\gamma _{ij}^{(n-ad)}=\frac{\left| J_{ij}\right| ^{2}}{\hbar }\sqrt{\frac{%
\pi }{\left( W_{i}+W_{j}\right) T}}\exp \left[ -\frac{\left( \Delta
+2W_{i}\right) ^{2}}{4\left( W_{i}+W_{j}\right) T}\right] .  \label{gamma4}
\end{equation}
which coincides with Eq.(\ref{gamma}). Otherwise, in the adiabatic limit
(see \cite{Emin} for details),

\begin{equation}
\gamma _{ij}^{(ad)}=\frac{1}{2\pi }\sqrt{\frac{W_{i}\Omega
_{i}^{2}+W_{j}\Omega _{j}^{2}}{W_{i}+W_{j}}}\exp \left\{ -\left[ \left(
\Delta +2W_{i}\right) ^{2}/4\left( W_{i}+W_{j}\right) -J_{ij}\right]
/T\right\}  \label{gamma5}
\end{equation}

When the vibrational and coupling constants of the initial and final sites
are the same, $W_{i}=W_{k}=W$, $\Omega _{i}=\Omega _{j}=\Omega $, the above
expressions for the hopping rate coincide with these obtained in Ref. \cite
{Emin}. An expression for the hopping rate, which is similar to Eq.(\ref
{gamma4}) but with a different pre-exponential factor, can be obtained for
the bound magnetic polarons by using a procedure described in Ref. \cite
{PF2000}. (This factor properly takes into account the vector nature of the
order parameter (magnetization) and can be responsible for the giant \textit{%
positive }hopping magnetoresistance that may be observed at low magnetic
fields \cite{PRL99}. However, at high magnetic fields it coincides with that
from Eq. (\ref{gamma4}).)


\end{document}